# Women Worry, Men Adopt? Gendered Risk Perceptions and Generative AI Adoption

Fabian Stephany [a,b,c] & Jedrzej Duszynski [a],

a) Oxford Internet Institute, University of Oxford, UK, b) Institute for New Economic Thinking, Oxford Martin School, c) Bruegel, Brussels, Belgium.

fabian.stephany@oii.ox.ac.uk

## Abstract

Generative artificial intelligence (GenAI) is spreading rapidly across work and daily life, yet adoption remains uneven. Men use GenAI more frequently than women, potentially widening inequalities in productivity, skills, and career opportunities. Existing research has largely explained this gap through differences in access, digital skills, and confidence. We argue that these explanations are incomplete: gender differences in GenAI adoption may also reflect how women and men evaluate AI's societal risks. Using two waves (2023–2024) of the nationally representative UK Public Attitudes to Data and AI Tracker ($N \approx 9{,}172$), we combine descriptive analyses with gender-specific, age-stratified random forest models and a parametric score-matching analysis of repeated cross-sections. We first show that men report substantially higher levels of frequent personal GenAI use than women. We then show that this gap is especially pronounced among respondents who express concerns about AI's societal consequences, particularly its effects on mental health and the environment. Intersectional analyses show that the largest disparities arise among younger, digitally fluent individuals with high societal risk concerns, where gender gaps in personal use exceed 45 percentage points. Across predictive models, perceived societal risk has greater predictive relevance for women's adoption than for men's and ranks among the strongest predictors of women's GenAI use. Finally, in score-matched comparisons, higher optimism about AI's societal impact is associated with larger increases in women's uptake, narrowing the gender gap. We interpret these findings as indication that unresolved AI harms may contribute to unequal access to GenAI's productivity, learning, and career benefits. The findings point to societal risk perception as an important behavioural pathway underlying digital inequality in the AI era.

**Acknowledgements:** The authors are grateful for the generous support of the Oxford Internet Institute's Research Programme on AI and Work, which is funded by the Dieter Schwarz Stiftung gGmbH. We also thank the Digital Gender Divide Research Group at Nuffield College for valuable feedback. We are particularly grateful to David Sutcliffe for excellent copy-editing and careful reviewing of the manuscript.

# Introduction

AI-driven technologies have been adopted more quickly than personal computers or the Internet, transforming both everyday- and work life, offering potential gains in productivity, efficiency, and task performance (Noy and Zhang, 2023; Babina et al., 2024; Brynjolfsson, Li and Raymond, 2025). Yet the benefits of generative AI are not distributed equally. Men are consistently more likely than women to adopt and frequently use these tools across countries, sectors, and occupations (Aldasoro et al., 2024; Otis et al., 2024). Existing research has largely explained this gender gap in terms of unequal access, digital skills, occupational exposure, and confidence. We argue that these explanations are important but incomplete. Gender differences in GenAI adoption may also reflect differences in how women and men evaluate the societal risks associated with AI.

This distinction matters because concerns about AI should not be treated simply as barriers to adoption or as individual reluctance to use new technology. Generative AI systems raise documented concerns around privacy, misinformation, bias, environmental impact, labour-market disruption, mental health, and broader social harms. If women are more attentive to these risks, lower adoption need not indicate irrational anxiety, lack of ability, or insufficient openness to innovation. It may instead reflect a reasonable response to unresolved harms. Our central contribution is therefore to examine whether societal risk perception is associated with unequal GenAI uptake.

Using two waves (2023–2024) of the nationally representative UK Public Attitudes to Data and AI Tracker, we show that men are substantially more likely than women to use GenAI frequently for personal purposes (20.9% versus 15.5%). The gender gap is especially wide among respondents who express concerns about AI's broader social consequences, particularly its effects on mental health and the environment. The largest disparities arise among younger, digitally fluent individuals with high societal risk concerns, where gender gaps in personal use exceed 45 percentage points. Across gender-specific predictive models, a composite measure of societal risk perception has greater predictive relevance for women's adoption than men's and ranks among the strongest predictors of women's use. In a parametric score-matching analysis of repeated cross-sections, higher optimism about AI's societal impact is associated with larger increases in women's adoption, narrowing the divide. These findings suggest that societal risk perception is an important behavioural pathway through which unresolved AI harms may shape unequal uptake.

The implication is not simply that women should be encouraged to adopt GenAI more quickly. If lower adoption partly reflects concern about real social and ethical harms, then closing the gender gap requires more than training, access, or confidence-building. It also requires reducing the risks that make hesitation a reasonable response. Safeguards, transparency, accountability, privacy protection, and bias mitigation are therefore not separate from the question of equitable adoption. They are central to ensuring that the benefits of GenAI can be shared broadly without asking more risk-sensitive groups to discount legitimate concerns.

# Background

## From Digital Divides to Unequal GenAI Adoption

The gender gap in generative AI adoption forms part of a longer history of unequal technological diffusion. Research on the digital divide has shown that inequalities in technology use persist even when basic access improves. Attewell (2001) distinguishes between primary digital divides, concerning access to technology, and secondary digital divides, concerning the skills, confidence, autonomy, and social conditions that shape how technology is used. As access gaps narrow, these secondary divides become more visible, revealing how technology use reflects broader socioeconomic inequalities (Cullen, 2001; van Dijk and Hacker, 2003; van Dijk, 2005, 2006, 2020; Joiner, Stewart and Beaney, 2015). Across many digital technologies, men have tended to adopt earlier and more frequently than women, even where differences in ability or performance are small or absent (Cooper, 2006; Galyani Moghaddam, 2010; Orji, 2010; Goswami and Dutta, 2016; Mariscal et al., 2019; BenYishay et al., 2020; Acilar and Sæbø, 2021).

These gaps matter because technology adoption is not only a matter of individual preference. Digital engagement is linked to skill development, workplace agency, labour-market participation, and earnings. Prior work has connected unequal access to and use of digital technologies with women's labour-market disadvantage and with the persistence of the gender wage gap (Sinha, 2018; Yufei, Xiangquan and Wenxin, 2018; Ma, 2022; Gao and Liu, 2023). Some scholars therefore frame the gender digital divide not simply as an efficiency problem, but as an issue of empowerment, equality, and human rights (Peacock, 2019; Treuthart, 2019; Acilar and Sæbø, 2021).

Emerging evidence suggests that generative AI reproduces this pattern. Men are more likely than women to use GenAI tools, both in general and in work-related contexts. Blandin, Bick and Deming (2024), using a nationally representative sample of more than 60,000 Americans aged 18–64, report an 8.45 percentage-point adoption gap in favour of men. Among Swedish university students, Stöhr, Ou and Malmström (2024) find that men report higher use, more positive attitudes, and fewer concerns regarding AI chatbots. Aldasoro et al. (2024), using a nationally representative sample of U.S. households, find that 50% of men had used generative AI compared with 37% of women. Similar patterns appear in workplace settings: Humlum and Vestergaard (2025) find that women in Denmark are 16 percentage points less likely than men to have used ChatGPT for work. Reviewing evidence across countries, sectors, and occupations, Otis et al. (2024) conclude that the GenAI gender gap appears to be widespread.

The consequences of unequal GenAI adoption may be especially important because the technology is diffusing during a period in which new workplace norms, productivity expectations, and AI-related skills are still forming. Experimental and firm-level evidence suggests that AI can increase productivity, efficiency, and task performance in some settings (Noy and Zhang, 2023; Czarnitzki, Fernández and Rammer, 2023; Babina et al., 2024; Brynjolfsson, Li and Raymond, 2025). Recent work also links AI skills to improved job prospects, higher pay, and broader career benefits (Bone, González Ehlinger and Stephany, 2025; Bloom et al., 2025; Castaneda, Bone and Stephany, 2025; Stephany, Teutloff and Leone, 2026). If men adopt GenAI earlier or more intensively, they may accumulate advantages in experimentation, skill acquisition, productivity, and workplace visibility. Unequal use may also shape the development of AI systems themselves. Because generative AI systems are embedded in feedback loops of prompting, correction, evaluation, and deployment, underrepresentation among users may mean that some groups have less influence over how these systems evolve (Cao, Koning and Nanda, 2024; Otis et al., 2024). Existing evidence on algorithmic bias and representational harm suggests that such feedback loops can reproduce or amplify social inequalities (Koenecke et al., 2020; Guilbeault et al., 2024).

## Why Skills and Exposure Are Not Enough

The existing literature explains gender gaps in technology adoption largely through differences in digital skills, confidence, occupational exposure, and socioeconomic position. These explanations are important, but they are incomplete. They help explain why some groups may have fewer opportunities to use new technologies, but they do not fully explain why individuals who are aware of GenAI, exposed to it, or capable of using it may still hesitate to adopt it.

Digital fluency is one established explanation. Women remain underrepresented in STEM fields despite narrowing gaps in formal access to education and despite limited evidence of consistent gender differences in STEM performance at school level (Wang and Degol, 2017; Siddiq and Scherer, 2019; Bachmann and Hertweck, 2025). This underrepresentation is shaped by cultural norms, gendered stereotypes, and male-favouring learning environments that associate technical competence with masculinity and reduce women's sense of belonging in STEM fields (Thébaud and Charles, 2018; Yoshikawa, Kokubo and Wu, 2018; Banchefsky, Lewis and Ito, 2019; Deiglmayr, Stern and Schubert, 2019; Höhne and Zander, 2019; Chan, 2022). Women also report lower self-efficacy in STEM contexts, even where performance differences do not justify such confidence gaps (Ellis, Fosdick and Rasmussen, 2016; Cimpian, Kim and McDermott, 2020; Zander et al., 2020). These patterns appear to carry over into generative AI: women report lower self-assessed knowledge of AI tools, lower confidence in their ability to use them effectively, and a greater need for training (Aldasoro et al., 2024; Carvajal, Franco and Isaksson, 2024; Stöhr, Ou and Malmström, 2024; Cachero, Tomás and Pujol, 2025; Humlum and Vestergaard, 2025).

Occupational and socioeconomic factors also matter. Men remain overrepresented in STEM-related, technology-intensive, and innovation-oriented occupations, where exposure to emerging digital tools may be more frequent and more institutionally supported (Wang and Degol, 2017; Cimpian, Kim and McDermott, 2020; Young, Wajcman and Sprejer, 2023; Beroíza-Valenzuela and Salas-Guzmán, 2024). Although new technological fields are sometimes imagined as opportunities to build more equitable industries from the start, this has rarely been realised in practice (Berman and Bourne, 2015). The emerging professions of AI and data science reflect the gendered history of computing and continue to reproduce patterns of exclusion and unequal participation (Young, Wajcman and Sprejer, 2023). At the same time, women are disproportionately represented in occupations highly exposed to AI, especially cognitive and clerical roles, implying both substantial potential to benefit from AI tools and heightened vulnerability to disruption if adoption remains uneven (Cazzaniga et al., 2025).

However, skills, confidence, and occupational exposure do not exhaust the problem. Recent evidence suggests that overall education explains little of the GenAI adoption gap, while age and workplace seniority account for only part of the observed differences (Méndez-Suárez, Monfort and Hervas-Oliver, 2023; Aldasoro et al., 2024; Blandin, Bick and Deming, 2024; Gupta, 2024; Humlum and Vestergaard, 2025). Moreover, treating lower adoption primarily as a skills or confidence deficit risks missing a distinctive feature of generative AI: adopting these systems requires users to engage with a technology whose social risks remain highly visible and unresolved. GenAI adoption is therefore not only a matter of capability or access. It also depends on whether users judge the technology to be trustworthy, legitimate, and safe enough to use.

## AI Harms and the Logic of Risk-Sensitive Adoption

Generative AI differs from many earlier workplace technologies because its diffusion has coincided with widespread public concern about social and ethical harms. These concerns should not be treated simply as individual anxiety or resistance to innovation. A growing literature documents concrete risks associated with large language models and algorithmic systems, including bias, representational harm, discrimination, privacy leakage, misinformation, environmental costs, labour-market disruption, and uneven quality of service across social groups. Mittelstadt et al. (2016) map the ethical concerns raised by algorithmic systems, including opacity, bias, accountability, and the difficulty of assigning responsibility for automated decisions. Bender et al. (2021) highlight risks related to scale, training data, environmental costs, and the reproduction of harmful language patterns. Weidinger et al. (2022) develop a taxonomy of language-model risks spanning discrimination, exclusion, information hazards, misinformation, malicious uses, and human-computer interaction harms. Shelby et al. (2023) similarly frame algorithmic harm as sociotechnical, distinguishing representational, allocative, quality-of-service, interpersonal, and societal harms.

This matters for how we interpret gendered adoption patterns. If women express greater concern about GenAI, this should not be read as irrational fear, technological disengagement, or excessive risk aversion. Research on digital risks and privacy suggests that women often report greater concern about online harms and data practices. Enock et al. (2024), for example, find that women are more fearful of experiencing online harms and less comfortable engaging in some online behaviours, despite similar exposure. Several studies also show that women tend to report higher privacy concerns across digital applications (Tifferet, 2019; Ho et al., 2022; Chen et al., 2023). Some work links gender gaps in technology adoption to differences in risk aversion, with women more deterred by perceived risks (Kanwal et al., 2021; Alonso et al., 2023). At the same time, the evidence on how gendered privacy concerns and risk attitudes translate into digital adoption is mixed (Gillam and Waite, 2021; McGill and Thompson, 2021; Lyons, Liu and Fang, 2024). Recent work on AI anxiety further suggests that women may report lower positive attitudes toward AI and lower levels of use (Russo et al., 2025).

Our argument builds on this literature but shifts the interpretation. Rather than treating risk perception as a psychological barrier to be overcome, we understand it as a potentially reasonable response to documented social harms. GenAI tools often require users to enter personal, professional, or sensitive information into systems whose data practices and downstream uses may be unclear. They may produce biased, inaccurate, or misleading outputs, while also raising concerns about labour displacement, environmental costs, and broader societal effects. In this context, concern about AI is not external to adoption. It is one of the conditions under which adoption decisions are made.

This risk-sensitive perspective is also supported by research on moral attitudes, social concern, and technology acceptance. Women have been found, on average, to report stronger social compassion, traditional moral concerns, and support for equity-oriented values (Pratto, Stallworth and Sidanius, 1997; Eagly et al., 2004). Moral and social concerns also shape technology acceptance beyond narrow self-interest. Kodapanakkal et al. (2020) show that acceptance of big data technologies is influenced not only by individual benefit but also by broader social concerns around privacy. Similar dynamics have been observed in the acceptance of smart grid and sustainable energy technologies, where perceived social benefits influence public acceptance (Huijts, Molin and Steg, 2012; Toft, Schuitema and Thøgersen, 2014). Gupta et al. (2015) likewise find that ethical concerns influence the acceptance of nanotechnology. In the GenAI context, studies of education suggest that women are more likely to perceive AI use in coursework or assignments as unethical, as facilitating cheating or plagiarism, or as contributing to misinformation (Carvajal, Franco and Isaksson, 2024; Stöhr, Ou and Malmström, 2024; Ravšelj et al., 2025).

These findings suggest that gendered GenAI adoption may partly reflect differences in how users weigh private benefits against social and ethical risks. Men's higher adoption and lower reported concern should not be treated as the normative benchmark. Nor should women's lower adoption be interpreted simply as a deficit in confidence, skill, or willingness to innovate. A central possibility is that women are more responsive to unresolved harms associated with GenAI, and that this responsiveness makes adoption less attractive even when the potential individual benefits are recognised.

## The Argument: Risk Perception as a Source of Unequal Benefit

The central claim of this study is that societal risk perception may be an important behavioural pathway linking AI harms to unequal adoption. Existing accounts of the GenAI gender gap rightly emphasise skills, confidence, exposure, and occupational position. We argue that these factors should be complemented by a risk-sensitive account. Because GenAI is associated with real and widely discussed societal harms, adoption decisions may depend not only on whether individuals can use the technology, but also on whether they judge its use to be acceptable, safe, and socially legitimate.

This reframes the gender gap in GenAI adoption. The problem is not that women are insufficiently enthusiastic about AI, nor that they should simply temper their risk concerns in order to access the benefits of adoption. Instead, unresolved risks may themselves generate inequality. If groups that are more attentive to AI's social and ethical harms are also less likely to adopt GenAI, then they may receive fewer of the productivity, learning, and career benefits associated with use. Unequal adoption then becomes partly a governance problem: unsafe, opaque, or insufficiently accountable AI systems may discourage participation among those who are more sensitive to their harms.

This perspective leads to three empirical expectations. First, gender gaps in GenAI use should be wider among respondents who express concern about AI's societal consequences, particularly in domains such as mental health, environmental impact, data privacy, and labour-market risk. Second, perceived societal risk should contribute to the prediction of GenAI use alongside conventional predictors such as age, education, occupation, and digital literacy, with greater predictive relevance for women than for men. Third, in repeated cross-sectional score-matching comparisons, higher optimism about AI's societal impact should be associated with greater GenAI uptake among otherwise comparable respondents, especially among women. Importantly, these expectations do not imply that lower risk perception is preferable. They imply that reducing the gender gap in GenAI adoption requires more than training, encouragement, or access. It requires addressing the harms that make hesitation a reasonable response in the first place.

## Data and Methods

### Data

We draw on two waves (2023 and 2024) of the UK Public Attitudes to Data and AI Tracker, a large-scale survey commissioned by the UK Department for Science, Innovation & Technology. Each wave contains responses from over 4,000 UK adults (4225 in 2023, 4947 in 2024), yielding a combined sample of 9,172 respondents (See Appendix Table A1. Sample Characteristics). Of those, approximately 220 respondents from each wave provided invalid responses to questions about their AI chatbot usage. The survey captures detailed demographic, socioeconomic, behavioural, and attitudinal information relating to data-driven technologies and artificial intelligence, making it well suited to studying gendered patterns of generative AI adoption.

The tracker is designed to be nationally representative of the adult UK population. In line with the survey design, analyses are conducted using the available survey weights where population-level descriptive estimates are reported. This is especially relevant for baseline levels of adoption and gender gaps in frequent use. The data provide repeated cross-sections rather than a true panel, such that the same individuals are not followed over time. To examine temporal shifts in attitudes and behaviour, we therefore complement cross-sectional analyses with a parametric score matching that pairs demographically comparable individuals across waves.

Our empirical focus is on generative AI adoption in both personal and work-related settings, and on whether gender differences in adoption are associated with differences in the perceived societal risks of AI. The analysis proceeds in three stages: descriptive analysis of gender gaps across subgroups, predictive modelling of GenAI use using age-stratified gender-specific machine-learning models, and a parametric score-matching analysis of changing literacy and risk perceptions across waves.

## Measures

**Generative AI usage.** The main outcome variable is self-reported frequency of generative AI use in personal and work-related contexts. In the main analyses, we define frequent use as use at least once per week. Although the survey includes a higher-frequency category, "multiple times a week," only around 7% of respondents select it, leaving small subgroup samples. Combining this category with "at least once a week" yields a frequent-use group of approximately 17% and preserves sufficient sample size for gender, age, and risk-perception comparisons. We note that this threshold reflects the early diffusion period covered by the 2023–2024 survey waves; what counted as frequent use then may be relatively modest by 2026. Personal and work-related use are analysed separately in descriptive analyses, while the predictive models focus on frequent personal use.

**Digital literacy.** Digital literacy is measured using respondents' self-reported understanding of AI and their ability to explain what AI is and how it works. The survey includes ordinal measures that capture varying levels of familiarity, ranging from limited understanding to being able to explain AI in detail. These indicators are used as proxies for perceived digital fluency and confidence with AI technologies.

**Risk perception.** Our main explanatory construct is a composite risk perception index intended to capture broader concern about the societal consequences of AI. The index is constructed from four binary indicators recording whether respondents perceive AI as posing risks in relation to: mental health, climate and environmental impact, data privacy, and the labour market. We average these four items to create a summary measure ranging from 0 to 1, where higher values indicate greater concern about AI's societal risks. This index is designed to capture other-oriented apprehension about AI rather than only individual-level fears about personal use.

**Demographic and socioeconomic covariates.** We use respondents' gender, age, education, and occupation / social grade as key background variables. Age is also used to define three broad life-stage groups for the predictive analyses: 18–35, 36–50, and 51+. These stratifications allow us to assess whether the predictors of GenAI adoption differ across generations and between men and women.

## Descriptive analysis

We begin by examining gender differences in frequent GenAI use across the full sample and across a wide range of demographic, behavioural, and attitudinal subgroups. This stage identifies the characteristics associated with wider or narrower gender gaps in both personal and work-related use. We also examine intersectional combinations of characteristics to identify cases in which gender disparities are amplified or attenuated. These descriptive analyses are based primarily on the 2024 wave, which provides the most recent snapshot of GenAI uptake.

### Predictive modelling

To assess the relative importance of risk perceptions alongside more established predictors of technology adoption, we estimate gender-specific random forest models stratified by age group based on the 2024 survey wave. Separate models are fit for women and men within each of the three age bands (18–35, 36–50, 51+). The outcome is frequent personal GenAI use. Predictors include digital literacy, education, occupation, and the composite risk perception index.

We use random forests because they allow for nonlinearities and higher-order interactions without requiring strong parametric assumptions, making them well suited to a setting where predictors may operate differently across demographic groups. Model performance is evaluated using the area under the receiver operating characteristic curve (AUC).

To compare the salience of different predictors, we report feature importance measures from each model. These scores indicate the relative contribution of each variable to predictive performance and are used to assess whether AI-related risk perceptions matter as much as, or more than, conventional demographic and skill-based predictors. The modelling exercise is intended as a transparent assessment of predictive relevance rather than as a causal identification strategy.

### Parametric score-matching design

Because the survey consists of repeated cross-sections rather than a panel, we cannot observe within-person changes in AI attitudes, literacy, or use. To examine whether shifts in literacy and societal AI optimism are associated with changes in GenAI adoption across waves, we therefore implement a parametric score-matching design. Respondents from the 2023 wave are matched to statistically comparable respondents from the 2024 wave using stable observed characteristics, including age, gender, education, and occupation / socioeconomic status. The matching score is estimated parametrically, and respondents are compared within matched strata to reduce observable differences between the two waves.

We use this framework to construct two matched comparisons: one comparing respondents with higher versus unchanged digital literacy, and one comparing respondents with higher versus unchanged optimism about AI's societal impact. For each comparison, we estimate the share of matched respondents whose personal GenAI use is higher in 2024 than among comparable respondents in 2023. The matching procedure is repeated 100 times, and estimates are summarised using means and 95% confidence intervals across iterations.

This design should not be interpreted as identifying causal effects. It does not follow the same individuals over time, and it cannot account for unobserved differences between matched respondents. Instead, it provides a structured repeated-cross-sectional comparison showing whether higher literacy or higher societal AI optimism is systematically associated with higher GenAI adoption among otherwise similar respondents. We therefore interpret the results as descriptive evidence consistent with a behavioural mechanism linking societal risk perception to adoption, rather than as causal estimates.

### Score-matched comparisons of changing attitudes and skills

Across the full sample, higher digital literacy and higher societal AI optimism are associated with greater GenAI uptake for both men and women. Among younger adults, however, the pattern differs by gender. Higher digital literacy is associated with a larger increase in GenAI uptake among men than among women, suggesting that capability-building alone may not close the gender gap. By contrast, higher optimism about AI's societal impact is associated with a larger increase in women's adoption, narrowing the gender divide. These results should be interpreted cautiously: they do not show that changing attitudes would mechanically cause women to adopt GenAI at higher rates. Rather, they suggest that perceived societal risk is closely linked to women's adoption decisions, even among respondents who are otherwise comparable on observed demographic and socioeconomic characteristics.

## Results

### Gender Differences in Generative AI Use and Risk Perceptions

We analyse data from the 2023 and 2024 waves of the UK Public Attitudes to Data and AI Tracker (Department for Science, Innovation & Technology; $N \approx$ 9,200). The survey measures both the frequency of generative AI (GenAI) use and respondents' concerns about potential societal risks associated with AI. Across the full sample, 15.5% of women and 20.9% of men report using GenAI tools frequently—defined as at least once per week—for personal purposes. This corresponds to a gender gap of 5.3 percentage points (Figure 1).

The gender gap varies across demographic groups. Among respondents aged 18–24, frequent personal use is reported by 22.0% of women and 35.9% of men, corresponding to a gap of 13.9 percentage points. Among those aged 25–34, the gap is 10.9 percentage points, and among those aged 35–44 it is 10.2 percentage points. The gap narrows in older age groups. Differences also emerge across respondents expressing specific AI-related concerns. Among respondents who report concern about AI's environmental impacts, the gender gap in frequent personal use is 9.3 percentage points. Among those concerned about AI's effects on mental health, the gap increases to 16.8 percentage points.

Similar patterns appear across additional risk domains, including concerns related to data privacy, data sharing, and labour-market impacts (Table A2&3). Across these groups, the share of frequent users is consistently lower among women than among men. The gap narrows for respondents reporting overall personal and societal optimism around AI, both at 4.7 percentage points. More details about the adoption levels for men and women for a wider range of respondent characteristics and perceptions in both the personal and work context can be found in Appendix Table A2 and A3, respectively.

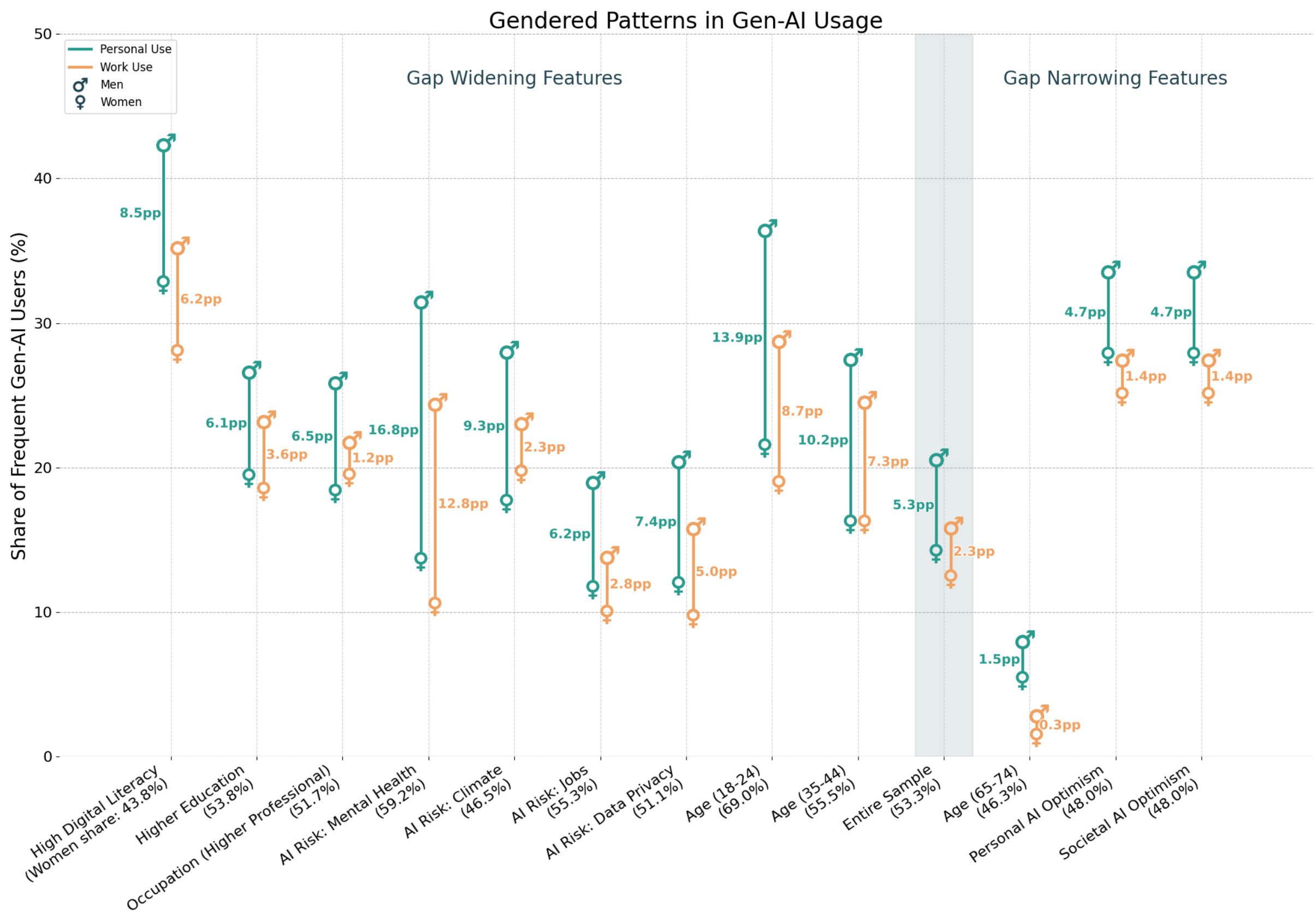

**Figure 1: Gendered patterns in GenAI usage.**
Gender gaps in frequent GenAI use across demographic, behavioural, and attitudinal groups show that gaps are wider among respondents concerned about AI's mental-health, environmental, privacy, and labour-market consequences.

## Intersectional Patterns in the Gender Gap

Figure 2 examines how the gender gap in GenAI use varies across selected pairwise combinations of demographic, behavioural, and attitudinal characteristics. The full set of comparisons is reported in Appendix Figure A2.

Across most intersections, gender differences are smaller for professional use than for personal use. However, certain combinations of factors produce substantially larger disparities. The largest gap—45.7 percentage points—appears for personal use among respondents who both perceive AI as a risk to mental health and report high levels of AI literacy. In the work context, the same group exhibits a smaller but still large gap of 29.6 percentage points. The second largest gap is observed for those concerned with AI's impact on mental health and data privacy, at 38.9 percentage points in personal use. Again, this gap is relatively smaller in the work context, at 19.1 percentage points.

Mental-health concerns amplify gender differences across many demographic groups, particularly when combined with younger age or higher digital literacy. By contrast, when mental-health concerns are paired with concerns about AI's labour-market consequences, the gender gap becomes considerably smaller (6.0 and 3.6 percentage points for personal and work use, respectively; see Appendix Figure A2).

Privacy concerns also interact with digital fluency. Among respondents who report both data-privacy concerns and high AI literacy, the gender gap in work-related use (25.0 percentage points) exceeds that observed for personal use (19.8 percentage points), one of the few cases where the pattern reverses.

Intersectional Gendered Patterns in Gen-AI Usage

Personal Usage

| | AI Risk: Mental Health | AI Risk: Climate | AI Risk: Data Privacy | High Digital Literacy | Higher Education | Age 18–24 |
|---|---|---|---|---|---|---|
| AI Risk: Mental Health | 19.5 | | 38.9 | 48.9 | 20.6 | 30.2 |
| AI Risk: Climate | | 9.1 | 22.4 | 2.7 | 3.8 | -2.4 |
| AI Risk: Data Privacy | 38.9 | 22.4 | 9.2 | 19.8 | 9.6 | 14.9 |
| High Digital Literacy | 48.9 | 2.7 | 19.8 | 7.7 | 9.9 | 5.9 |
| Higher Education | 20.6 | 3.8 | 9.6 | 9.9 | 5.9 | 15.5 |
| Age 18–24 | 30.2 | -2.4 | 14.9 | 5.9 | 15.5 | 9.4 |

Work Usage

| | AI Risk: Mental Health | AI Risk: Climate | AI Risk: Data Privacy | High Digital Literacy | Higher Education | Age 18–24 |
|---|---|---|---|---|---|---|
| AI Risk: Mental Health | 12.0 | | 19.1 | 28.8 | 18.2 | 24.6 |
| AI Risk: Climate | | 2.5 | 14.1 | 1.8 | -6.5 | 11.0 |
| AI Risk: Data Privacy | 19.1 | 14.1 | 6.1 | 25.0 | 6.6 | 12.6 |
| High Digital Literacy | 28.8 | 1.8 | 25.0 | 5.8 | 2.7 | 7.0 |
| Higher Education | 18.2 | -6.5 | 6.6 | 2.7 | 2.1 | 5.5 |
| Age 18–24 | 24.6 | 11.0 | 12.6 | 7.0 | 5.5 | 7.9 |

Note: Empty boxes indicate <50 observations in the intersection sample.

**Figure 2: Gendered and intersectional patterns in GenAI usage.**
Gender AI gap for various subsamples of the population. Risk perceptions interact with age, education, and digital literacy to produce markedly larger gender disparities in some subgroups.

## Relative Importance of Risk Perceptions

To evaluate the importance of risk perceptions relative to demographic and skills-based predictors, we construct a composite AI risk perception index based on four binary indicators capturing concern about AI's impact on mental health, climate change, data privacy, and labour markets. These indicators are averaged to produce a scale ranging from 0 to 1 (mean $\approx$ 0.20).

We estimate gender-specific random forest models predicting the frequency of personal GenAI use. Models are stratified by age group (18–35, 36–50, and 51+) and include predictors for digital literacy, education, occupation, and the composite risk index. This non-parametric modelling approach allows us to capture nonlinear relationships and assess the relative contribution of each predictor (Figure 3).

Across all age groups, risk perceptions contribute more strongly to model predictions for women than for men. Among young adults, risk perceptions rank second in importance for women but fifth for men. Among middle-aged adults, risk perceptions rank first for both men and women. Among older adults, they risk second for men and first for women.

Model performance is stable across age strata (AUC $\approx$ 0.61–0.70). Risk perceptions account for 9–20% of total feature importance, indicating that attitudinal factors contribute meaningfully to the prediction of AI adoption alongside digital skills and demographic characteristics.

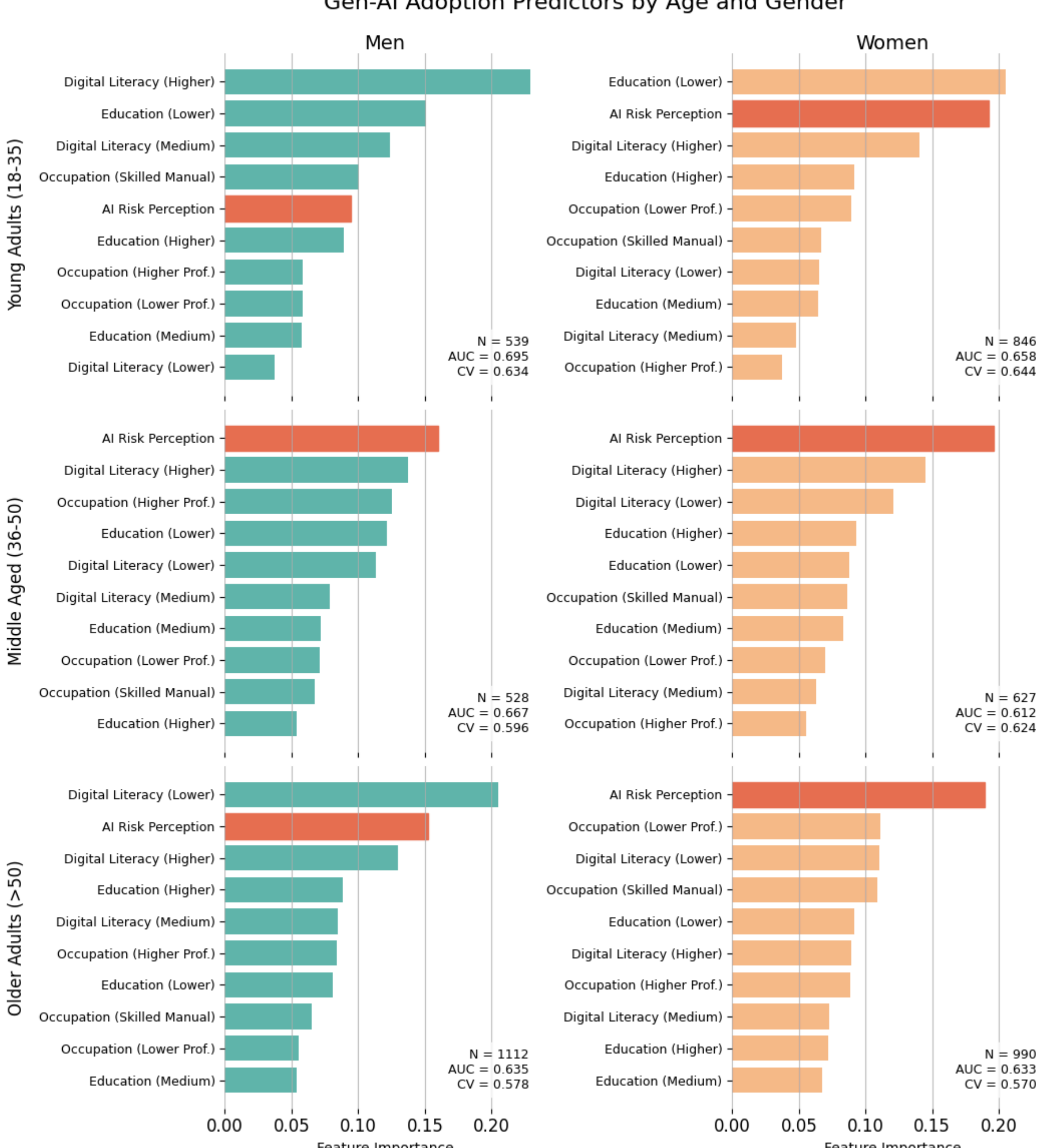

**Figure 3. Predictors of frequent GenAI adoption by age and gender.**
Gender-specific random-forest models (stratified by age groups) estimate the relative importance of digital skills, education, occupation, and a composite AI-risk-perception index in predicting frequent personal GenAI use (The feature importance scores tell how much a feature influences model predictions, with higher scores meaning greater importance with a maximum score of one). Across all life stages, risk perceptions have greater predictive relevance for women than for men—ranking among the top predictors for women in every age group, and surpassing demographic and skill factors. Model performance is stable across strata (AUC $\approx$ 0.61–0.70).

### Score-Matched Comparisons of Changing Attitudes and Skills

To assess whether differences in perceptions or skills are associated with differences in GenAI uptake across waves, we use a parametric score-matching analysis of repeated cross-sections. Because the same individuals are not observed in both waves, this analysis does not track within-person change. Instead, it compares respondents from 2023 and 2024 who are similar on observed demographic and socioeconomic characteristics but differ in digital literacy or societal AI optimism.

For each respondent in Wave 3, we match Wave 4 individuals with identical observable characteristics (age, gender, education, and occupation). This approach approximates a counterfactual comparison between individuals whose observable characteristics remain similar but whose skills or attitudes change between waves. We then compare changes in personal GenAI use among individuals who either improved their digital literacy or became more optimistic about AI's societal impact.

Figure 4 presents the parametric score matching approach, analytical approach, and results for young adults aged 18–35. The score-matched comparison results for the entire sample can be found in Appendix Figure A4.

Across the full sample, higher digital literacy and higher societal AI optimism are associated with higher GenAI uptake for both men and women (see Appendix Figure A4). This is not the case among younger adults. Higher digital literacy is associated with an insignificant increase in GenAI uptake for women (18%→31%) but a substantial and statistically significant increase for men (19%→46%). By contrast, higher optimism about AI's societal impact is associated with a larger and significant adoption increase among women (14%→34%), with a relatively smaller and insignificant change among men (22%→38%).

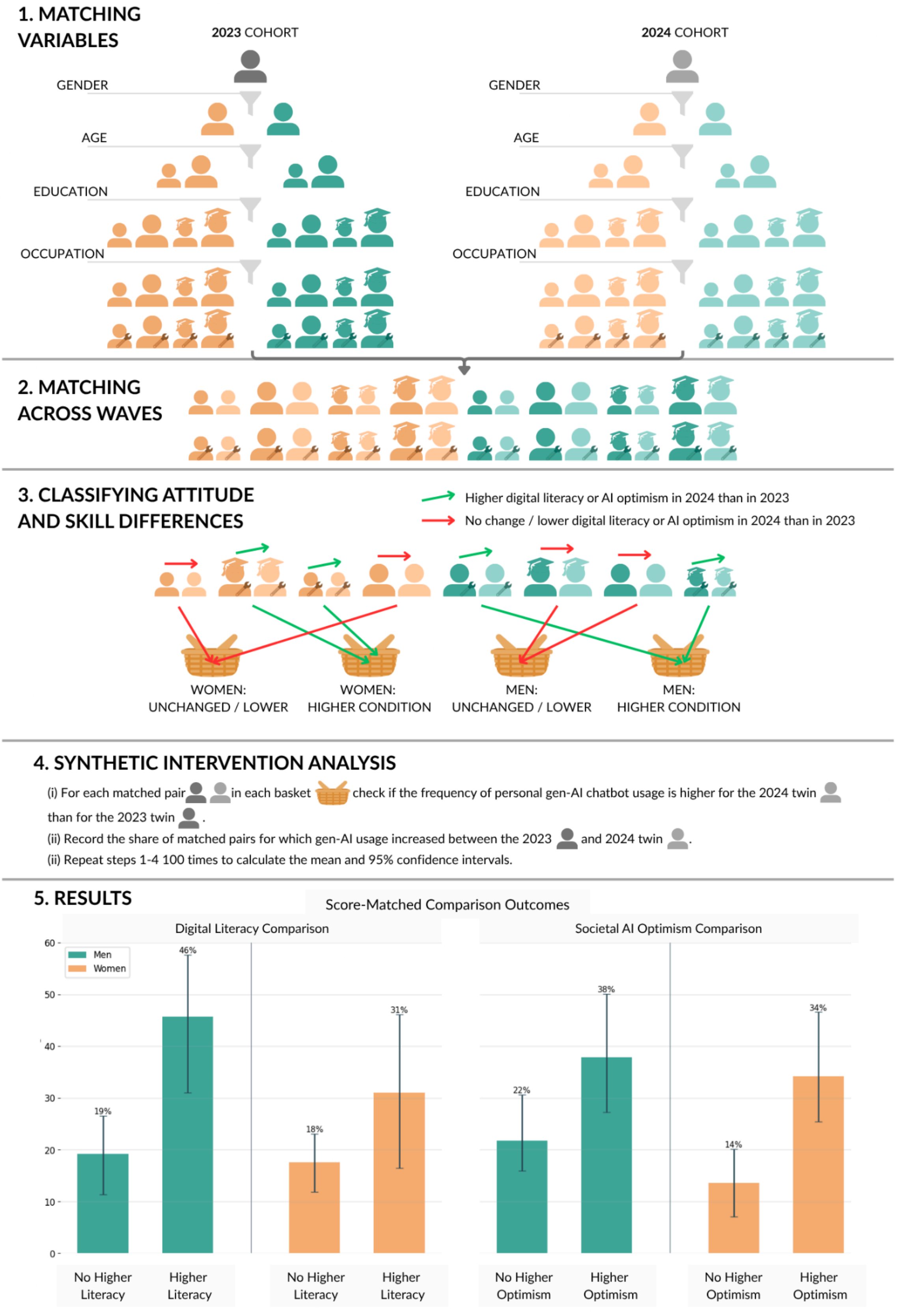

**Figure 4. Parametric score-matching design and matched differences in personal GenAI use.**

The figure illustrates the repeated cross-sectional matching approach used to compare GenAI uptake across the 2023 and 2024 survey waves. Respondents are matched across waves using observed demographic and socioeconomic characteristics, including gender, age, education, and occupation. Within these matched comparisons, respondents are classified according to whether the 2024 respondent reports higher digital literacy or higher societal AI optimism than the matched 2023 respondent. We then estimate the share of matched comparisons in which personal GenAI use is higher in 2024. The procedure is repeated 100 times to calculate mean matched differences and 95% confidence intervals. The bottom panels show results for young adults aged 18–35. Higher digital literacy is associated with greater GenAI uptake especially among men, while higher societal AI optimism is associated with larger increases among women.

## Discussion

### Findings

This study examined gender differences in generative AI adoption through the lens of perceived societal risk. Using UK survey data from 2023 and 2024, we find that men report higher levels of frequent GenAI use than women, and that this gap is not uniform across the population. Although age, education, occupation, and digital literacy remain important correlates of adoption, perceptions of AI's broader social consequences are also closely associated with gendered patterns of use. Across descriptive comparisons, predictive models, and score-matched comparisons, risk perceptions are more strongly linked to women's GenAI adoption than to men's. Concerns related to mental health, environmental impact, privacy, and labour-market disruption are associated with lower levels of GenAI use among women and with wider gender gaps in adoption.

These results are consistent with a risk-sensitive account of technology adoption. Much of the literature on digital inequality focuses on differences in access, capability, confidence, or economic incentives. Our findings suggest that GenAI adoption may also depend on how individuals evaluate the broader social consequences of the technology. Women's lower rates of adoption are especially pronounced where AI is perceived as threatening mental health, environmental sustainability, data privacy, or employment. In this sense, the gender gap in GenAI use may partly reflect differences in how societal externalities are weighed when individuals decide whether and how to engage with emerging technologies.

The findings contribute to the growing literature on unequal adoption of generative AI (Aldasoro et al., 2024; Blandin, Bick and Deming, 2024; Humlum and Vestergaard, 2025) by identifying societal risk perception as a behavioural pathway associated with adoption gaps. Previous work has shown that men tend to adopt generative AI tools earlier and more frequently across occupations and regions. Our results suggest that part of this difference may be linked to gendered differences in perceived societal risk. In the random-forest models, the composite risk-perception measure consistently ranks among the more important predictors of women's adoption, often alongside or above conventional predictors such as education or occupational status.

These patterns are particularly visible among younger respondents, who are also among the most frequent users overall. The gender gap therefore appears not only among groups with lower digital engagement, but also among cohorts most exposed to the rapid diffusion of generative AI. This suggests that adoption differences may emerge early in the life cycle of new technologies, while usage norms and workplace expectations are still forming. If these early differences persist, they may contribute to longer-term disparities in skill development, productivity, and career advancement as generative AI becomes embedded in everyday work, with recent evidence suggesting that unequal use may already be linked to gendered productivity differences (Tang et al., 2025).

The score-matched comparisons provide further evidence consistent with this interpretation. Higher digital literacy is associated with greater GenAI uptake for both men and women, but among younger adults this association is stronger for men, meaning that literacy-related differences may widen rather than close the gender gap. By contrast, higher optimism about AI's societal impact is associated with comparatively larger increases in women's usage, narrowing the gap in matched comparisons. These results should not be interpreted as causal effects, since the data consist of repeated cross-sections rather than a true panel. They nevertheless suggest that women's adoption is closely linked to how AI's societal consequences are evaluated, even among respondents who are similar on observed demographic and socioeconomic characteristics.

This interpretation aligns with a wider body of research showing that women often report stronger sensitivity to social and ethical risks associated with new technologies (Russo et al., 2025). Earlier studies of digital risk have often emphasized women's concerns about harms they may personally face online. Our findings point to a complementary dynamic: women's hesitation toward generative AI appears to be associated not only with self-oriented risk, but also with concern about harms imposed on others or on society more broadly. In that sense, lower adoption may partly reflect risk-sensitive evaluation rather than technological disengagement. Recognizing this distinction is important because it reframes part of the gender gap from a question of missing capacity to one of differing assessments of AI's legitimacy, safety, and social value.

## Implications

These results have implications for how policymakers and organisations approach unequal technological uptake. A common response to adoption gaps is to focus on training, digital literacy, and access to tools. Such interventions remain important, but our findings suggest that capability-building alone may be insufficient if adoption decisions are also shaped by concerns about AI's societal legitimacy and downstream harms. In particular, the results should not be read as implying that women should simply become more accepting of AI, or that lower risk perception among men represents the desirable benchmark. If hesitation partly reflects concern about documented risks, then interventions aimed solely at improving proficiency risk addressing only one part of the adoption gap.

A complementary approach would focus on reducing the risks that make hesitation reasonable. This includes strengthening privacy protections, improving accountability mechanisms, increasing transparency around the environmental and labour impacts of AI development, and mitigating well-documented risks related to bias, misinformation, and uneven quality of service. Addressing these issues would not only improve the technology itself, but may also increase trust among groups that are currently more hesitant to adopt it. In this sense, narrowing the gender gap is not only about changing user behaviour; it is also about making generative AI systems more trustworthy, accountable, and socially responsive.

More broadly, the results point to a potential mechanism through which inequality may compound over time. If men adopt generative AI more quickly while workplace norms and AI-related competencies are still emerging, early adoption advantages may become associated with longer-term disparities in productivity, visibility, and career progression. Similar dynamics have been documented in earlier waves of digital adoption, where differences in early engagement were linked to measurable labour-market outcomes. Given the central role generative AI is expected to play in future knowledge work, understanding these behavioural adoption patterns is critical, particularly in light of evidence that women are both highly exposed to AI-affected occupations and differentially vulnerable to its labour-market impacts (Cazzaniga et al., 2025).

## Limitations

The findings of this study should be considered in the context of several limitations. First, the analyses rely on self-reported measures of AI usage and digital literacy, which may diverge from actual behaviour or technical capability. Self-assessed literacy may in particular capture confidence or familiarity as much as substantive skill. Second, the data consist of repeated cross-sections rather than a true panel, meaning that we cannot observe within-person changes in AI attitudes, literacy, or use. The parametric score-matching analysis compares demographically similar respondents across waves, but it cannot account for unobserved differences between matched individuals and should not be interpreted as identifying causal effects. Third, the study focuses on the UK context, where the aggregate gender gap in generative AI adoption is smaller than has been reported in some international evidence. The magnitude and structure of the observed gaps may therefore vary across institutional, cultural, and labour-market settings.

Future research should investigate how risk perceptions are associated with generative AI adoption across different countries, sectors, and stages of technological diffusion. Combining survey data with behavioural usage traces, panel data, or experimental designs could help identify more precisely how societal risk perception relates to adoption decisions. It will also be important to examine whether individuals who hesitate to adopt AI tools differ not only in whether they use them, but also in how they use them—for example in the types of tasks delegated to AI, the prompting strategies employed, or the degree of trust placed in AI outputs.

## Conclusion

Generative AI is diffusing rapidly across workplaces, education, and everyday problem-solving. As with previous waves of digital technology, early patterns of adoption may shape who benefits most from the productivity gains and opportunities these tools enable. This study provides evidence that the emerging gender gap in generative AI use is associated not only with differences in skills, access, or socioeconomic position, but also with perceptions of AI's societal risks. Concerns related to mental health, environmental impact, data privacy, and employment are closely linked to gendered patterns of GenAI adoption.

Across descriptive analyses, predictive models, and parametric score-matching comparisons, we find consistent evidence that these risk perceptions are more strongly associated with adoption among women than among men. Women who express concerns about AI's broader social consequences are less likely to use generative AI tools frequently, and these differences are especially pronounced among younger and digitally literate groups. At the same time, greater optimism about AI's societal impact is associated with higher women's uptake in matched comparisons, suggesting that adoption is linked not only to capability, but also to how the technology's social consequences are evaluated.

Taken together, these results highlight a behavioural dimension of technological adoption that is often overlooked in discussions of digital inequality. Much of the policy debate around AI skills focuses on expanding technical capability and access to tools. While such interventions remain important, our findings suggest that adoption decisions are also associated with how individuals evaluate the broader social consequences of emerging technologies. In the case of generative AI, women appear more likely to weigh these societal considerations when deciding whether to engage with the technology.

Recognising this dynamic has implications for both research and policy. If adoption differences are partly rooted in perceptions of social and ethical risk, then reducing the gender gap requires not only improving digital capability, but also addressing the legitimacy and governance of AI systems themselves. Policies that strengthen transparency, accountability, environmental sustainability, privacy protection, and safeguards against bias may therefore matter not only for responsible AI development, but also for equitable adoption.

More broadly, the findings underscore how values and perceptions can shape the diffusion of new technologies. The point is not that women should become less risk-sensitive, or that higher adoption among men represents the ideal benchmark. Rather, unresolved AI harms may make hesitation a reasonable response, while also limiting access to the benefits of GenAI. As generative AI continues to transform knowledge work and digital interaction, ensuring that its benefits are broadly shared will require making these systems safer, more trustworthy, and more socially responsive.

## References

Acilar, A. and Sæbø, Ø. (2021) 'Towards understanding the gender digital divide: a systematic literature review', *Global Knowledge, Memory and Communication*, 72(3), pp. 233–249. Available at: https://doi.org/10.1108/GKMC-09-2021-0147.

Aldasoro, I. *et al.* (2024) 'The gen AI gender gap', *Economics Letters*, 241, p. 111814. Available at: https://doi.org/10.1016/j.econlet.2024.111814.

Alonso, S.L.N. *et al.* (2023) 'Gender gap in the ownership and use of cryptocurrencies: Empirical evidence from Spain', *Journal of Open Innovation: Technology, Market, and Complexity*, 9(3), p. 100103. Available at: https://doi.org/10.1016/j.joitmc.2023.100103.

Attewell, P. (2001) 'Comment: The First and Second Digital Divides', *Sociology of Education*, 74(3), pp. 252–259. Available at: https://doi.org/10.2307/2673277.

Babina, T. *et al.* (2024) 'Artificial intelligence, firm growth, and product innovation', *Journal of Financial Economics*, 151, p. 103745. Available at: https://doi.org/10.1016/j.jfineco.2023.103745.

Bachmann, R. and and Hertweck, F. (2025) 'The gender gap in digital literacy: a cohort analysis for Germany', *Applied Economics Letters*, 32(5), pp. 608–613. Available at: https://doi.org/10.1080/13504851.2023.2277685.

Banchefsky, S., Lewis, K.L. and Ito, T.A. (2019) 'The Role of Social and Ability Belonging in Men's and Women's pSTEM Persistence', *Frontiers in Psychology*, 10. Available at: https://doi.org/10.3389/fpsyg.2019.02386.

BenYishay, A. *et al.* (2020) 'Gender gaps in technology diffusion', *Journal of Development Economics*, 143, p. 102380. Available at: https://doi.org/10.1016/j.jdeveco.2019.102380.

Berman, F.D. and Bourne, P.E. (2015) 'Let's Make Gender Diversity in Data Science a Priority Right from the Start', *PLOS Biology*, 13(7), p. e1002206. Available at: https://doi.org/10.1371/journal.pbio.1002206.

Beroíza-Valenzuela, F. and Salas-Guzmán, N. (2024) 'STEM and gender gap: a systematic review in WoS, Scopus, and ERIC databases (2012–2022)', *Frontiers in Education*, 9. Available at: https://doi.org/10.3389/feduc.2024.1378640.

Blandin, A., Bick, A. and Deming, D. (2024) 'The Rapid Adoption of Generative AI'. Rochester, NY: Social Science Research Network. Available at: https://doi.org/10.2139/ssrn.4965142.

Bloom, D.E. *et al.* (2025) 'Artificial intelligence and the skill premium', *Finance Research Letters*, 81, p. 107401. Available at: https://doi.org/10.1016/j.frl.2025.107401.

Bone, M., González Ehlinger, E. and Stephany, F. (2025) 'Skills or degree? The rise of skill-based hiring for AI and green jobs', *Technological Forecasting and Social Change*, 214, p. 124042. Available at: https://doi.org/10.1016/j.techfore.2025.124042.

Brynjolfsson, E., Li, D. and Raymond, L. (2025) 'Generative AI at Work*', *The Quarterly Journal of Economics*, 140(2), pp. 889–942. Available at: https://doi.org/10.1093/qje/qjae044.

Cachero, C., Tomás, D. and Pujol, F.A. (2025) 'Gender Bias in Self-Perception of AI Knowledge, Impact, and Support among Higher Education Students: An Observational Study', *ACM Trans. Comput. Educ.*, 25(2), p. 15:1-15:26. Available at: https://doi.org/10.1145/3721295.

Cao, R., Koning, R. and Nanda, R. (2024) 'Sampling Bias in Entrepreneurial Experiments', *Management Science*, 70(10), pp. 7283–7307. Available at: https://doi.org/10.1287/mnsc.2021.01740.

Carvajal, D., Franco, C. and Isaksson, S. (2024) *Will Artificial Intelligence Get in the Way of Achieving Gender Equality?*, *122*. Working paper. Institutt for samfunnsøkonomi. Available at: https://openaccess.nhh.no/nhh-xmlui/handle/11250/3122396 (Accessed: 8 May 2025).

Cazzaniga, M. *et al.* (2025) 'A Gender Lens on Labor Market Exposure to AI', *AEA Papers and Proceedings*, 115, pp. 56–61. Available at: https://doi.org/10.1257/pandp.20251046.

Chan, R.C.H. (2022) 'A social cognitive perspective on gender disparities in self-efficacy, interest, and aspirations in science, technology, engineering, and mathematics (STEM): the influence of cultural and gender norms', *International Journal of STEM Education*, 9(1), p. 37. Available at: https://doi.org/10.1186/s40594-022-00352-0.

Chen, S. *et al.* (2023) 'The fintech gender gap', *Journal of Financial Intermediation*, 54, p. 101026. Available at: https://doi.org/10.1016/j.jfi.2023.101026.

Cimpian, J.R., Kim, T.H. and McDermott, Z.T. (2020) 'Understanding persistent gender gaps in STEM', *Science*, 368(6497), pp. 1317–1319. Available at: https://doi.org/10.1126/science.aba7377.

Cooper, J. (2006) 'The digital divide: the special case of gender', *Journal of Computer Assisted Learning*, 22(5), pp. 320–334. Available at: https://doi.org/10.1111/j.1365-2729.2006.00185.x.

Cullen, R. (2001) 'Addressing the digital divide', *Online Information Review*, 25(5), pp. 311–320. Available at: https://doi.org/10.1108/14684520110410517.

Czarnitzki, D., Fernández, G.P. and Rammer, C. (2023) 'Artificial intelligence and firm-level productivity', *Journal of Economic Behavior & Organization*, 211, pp. 188–205. Available at: https://doi.org/10.1016/j.jebo.2023.05.008.

Deiglmayr, A., Stern, E. and Schubert, R. (2019) 'Beliefs in "Brilliance" and Belonging Uncertainty in Male and Female STEM Students', *Frontiers in Psychology*, 10. Available at: https://doi.org/10.3389/fpsyg.2019.01114.

van Dijk, J. (2020) *The Digital Divide*. John Wiley & Sons.

van Dijk, J. and and Hacker, K. (2003) 'The Digital Divide as a Complex and Dynamic Phenomenon', *The Information Society*, 19(4), pp. 315–326. Available at: https://doi.org/10.1080/01972240309487.

van Dijk, J.A.G.M. (2005) *The Deepening Divide: Inequality in the Information Society*. SAGE Publications.

van Dijk, J.A.G.M. (2006) 'Digital divide research, achievements and shortcomings', *Poetics*, 34(4), pp. 221–235. Available at: https://doi.org/10.1016/j.poetic.2006.05.004.

Eagly, A.H. *et al.* (2004) 'Gender Gaps in Sociopolitical Attitudes: A Social Psychological Analysis', *Journal of Personality and Social Psychology*, 87(6), pp. 796–816. Available at: https://doi.org/10.1037/0022-3514.87.6.796.

Ellis, J., Fosdick, B.K. and Rasmussen, C. (2016) 'Women 1.5 Times More Likely to Leave STEM Pipeline after Calculus Compared to Men: Lack of Mathematical Confidence a Potential Culprit', *PLOS ONE*, 11(7), p. e0157447. Available at: https://doi.org/10.1371/journal.pone.0157447.

Enock, F.E. *et al.* (2024) 'Understanding gender differences in experiences and concerns surrounding online harms: A short report on a nationally representative survey of UK adults'. arXiv. Available at: https://doi.org/10.48550/arXiv.2402.00463.

Galyani Moghaddam, G. (2010) 'Information technology and gender gap: toward a global view', *The Electronic Library*, 28(5), pp. 722–733. Available at: https://doi.org/10.1108/02640471011081997.

Gao, J. and Liu, Y. (2023) 'Has Internet Usage Really Narrowed the Gender Wage Gap?: Evidence from Chinese General Social Survey Data', *Human Behavior and Emerging Technologies*, 2023(1), p. 7580041. Available at: https://doi.org/10.1155/2023/7580041.

Gillam, A.R. and Waite, A.M. (2021) 'Gender differences in predictors of technology threat avoidance', *Information & Computer Security*, 29(3), pp. 393–412. Available at: https://doi.org/10.1108/ICS-01-2020-0008.

Goswami, A. and Dutta, S. (2016) 'Gender Differences in Technology Usage—A Literature Review', *Open Journal of Business and Management*, 04(01), pp. 51–59. Available at: https://doi.org/10.4236/ojbm.2016.41006.

Guilbeault, D. *et al.* (2024) 'Online images amplify gender bias', *Nature*, 626(8001), pp. 1049–1055. Available at: https://doi.org/10.1038/s41586-024-07068-x.

Gupta, N., Fischer, A.R.H. and Frewer, L.J. (2015) 'Ethics, Risk and Benefits Associated with Different Applications of Nanotechnology: a Comparison of Expert and Consumer Perceptions of Drivers of Societal Acceptance', *NanoEthics*, 9(2), pp. 93–108. Available at: https://doi.org/10.1007/s11569-015-0222-5.

Gupta, V. (2024) 'An Empirical Evaluation of a Generative Artificial Intelligence Technology Adoption Model from Entrepreneurs' Perspectives', *Systems*, 12(3), p. 103. Available at: https://doi.org/10.3390/systems12030103.

Ho, M.-T. *et al.* (2022) 'Rethinking technological acceptance in the age of emotional AI: Surveying Gen Z (Zoomer) attitudes toward non-conscious data collection', *Technology in Society*, 70, p. 102011. Available at: https://doi.org/10.1016/j.techsoc.2022.102011.

Höhne, E. and Zander, L. (2019) ‘Sources of Male and Female Students’ Belonging Uncertainty in the Computer Sciences’, *Frontiers in Psychology*, 10. Available at: https://doi.org/10.3389/fpsyg.2019.01740.

Huijts, N.M.A., Molin, E.J.E. and Steg, L. (2012) ‘Psychological factors influencing sustainable energy technology acceptance: A review-based comprehensive framework’, *Renewable and Sustainable Energy Reviews*, 16(1), pp. 525–531. Available at: https://doi.org/10.1016/j.rser.2011.08.018.

Humlum, A. and Vestergaard, E. (2025) ‘The unequal adoption of ChatGPT exacerbates existing inequalities among workers’, *Proceedings of the National Academy of Sciences*, 122(1), p. e2414972121. Available at: https://doi.org/10.1073/pnas.2414972121.

Joiner, R., Stewart, C. and Beaney, C. (2015) ‘Gender Digital Divide’, *The Wiley Handbook of Psychology, Technology, and Society*. John Wiley & Sons, Ltd, pp. 74–88. Available at: https://doi.org/10.1002/9781118771952.ch4.

Kanwal, M. *et al.* (2021) ‘Systematic review of gender differences and similarities in online consumers’ shopping behavior’, *Journal of Consumer Marketing*, 39(1), pp. 29–43. Available at: https://doi.org/10.1108/JCM-01-2021-4356.

Kodapanakkal, R.I. *et al.* (2020) ‘Self-interest and data protection drive the adoption and moral acceptability of big data technologies: A conjoint analysis approach’, *Computers in Human Behavior*, 108, p. 106303. Available at: https://doi.org/10.1016/j.chb.2020.106303.

Koenecke, A. *et al.* (2020) ‘Racial disparities in automated speech recognition’, *Proceedings of the National Academy of Sciences*, 117(14), pp. 7684–7689. Available at: https://doi.org/10.1073/pnas.1915768117.

Lyons, A., Liu, F. and Fang, E.S. (2024) ‘Risk, Gender, and Digital Finance’. Rochester, NY: Social Science Research Network. Available at: https://doi.org/10.2139/ssrn.4914727.

Ma, X. (2022) ‘Internet use and gender wage gap: evidence from China’, *Journal for Labour Market Research*, 56(1), p. 15. Available at: https://doi.org/10.1186/s12651-022-00320-9.

Mariscal, J. *et al.* (2019) ‘Bridging the Gender Digital Gap’, *Economics*, 13(1), p. 20190009. Available at: https://doi.org/10.5018/economics-ejournal.ja.2019-9.

McGill, T. and Thompson, N. (2021) ‘Exploring potential gender differences in information security and privacy’, *Information & Computer Security*, 29(5), pp. 850–865. Available at: https://doi.org/10.1108/ICS-07-2020-0125.

Méndez-Suárez, M., Monfort, A. and Hervas-Oliver, J.-L. (2023) ‘Are you adopting artificial intelligence products? Social-demographic factors to explain customer acceptance’, *European Research on Management and Business Economics*, 29(3), p. 100223. Available at: https://doi.org/10.1016/j.iedeen.2023.100223.

Castaneda, A., Bone, M. and Stephany, F. (2025) ‘Beyond pay: AI skills reward more job benefits’. arXiv. Available at: https://doi.org/10.48550/arXiv.2507.20410.

Mittelstadt, B.D. *et al.* (2016) 'The ethics of algorithms: Mapping the debate', *Big Data & Society*, 3(2), p. 2053951716679679. Available at: https://doi.org/10.1177/2053951716679679.

Noy, S. and Zhang, W. (2023) 'Experimental evidence on the productivity effects of generative artificial intelligence', *Science*, 381(6654), pp. 187–192. Available at: https://doi.org/10.1126/science.adh2586.

Orji, R. (2010) 'Impact of Gender and Nationality on Acceptance of a Digital Library: An Empirical Validation of Nationality Based UTAUT Using SEM', *Journal of Emerging Trends in Computing and* ... [Preprint]. Available at: https://www.academia.edu/3233603/Impact_of_Gender_and_Nationality_on_Acceptance_of_a_Digital_Library_An_Empirical_Validation_of_Nationality_Based_UTAUT_Using_SEM (Accessed: 8 May 2025).

Otis, N.G. *et al.* (2024) 'Global Evidence on Gender Gaps and Generative AI'. Available at: https://doi.org/10.31219/osf.io/h6a7c.

Parveen, M. and Alkudsi, Y.M. (2024) 'Graduates' Perspectives on AI Integration: Implications for Skill Development and Career Readiness', *IJERI: International Journal of Educational Research and Innovation*, (22), pp. 1–17. Available at: https://doi.org/10.46661/ijeri.10651.

Peacock, A. (2019) *Human Rights and the Digital Divide*. London: Routledge. Available at: https://doi.org/10.4324/9781351046794.

Pratto, F., Stallworth, L.M. and Sidanius, J. (1997) 'The gender gap: Differences in political attitudes and social dominance orientation', *British Journal of Social Psychology*, 36(1), pp. 49–68. Available at: https://doi.org/10.1111/j.2044-8309.1997.tb01118.x.

Ravšelj, D. *et al.* (2025) 'Higher education students' perceptions of ChatGPT: A global study of early reactions', *PLOS ONE*, 20(2), p. e0315011. Available at: https://doi.org/10.1371/journal.pone.0315011.

Russo, C. *et al.* (2025) 'Gender differences in artificial intelligence: the role of artificial intelligence anxiety', *Frontiers in Psychology*, 16. Available at: https://doi.org/10.3389/fpsyg.2025.1559457.

Siddiq, F. and Scherer, R. (2019) 'Is there a gender gap? A meta-analysis of the gender differences in students' ICT literacy', *Educational Research Review*, 27, pp. 205–217. Available at: https://doi.org/10.1016/j.edurev.2019.03.007.

Sinha, S. (2018) 'Gender Digital Divide in India: Impacting Women's Participation in the Labour Market', in NILERD (ed.) *Reflecting on India's Development: Employment, Skill and Health*. Singapore: Springer, pp. 293–310. Available at: https://doi.org/10.1007/978-981-13-1414-8_14.

Stephany, F., Teutloff, O. and Leone, A. (2026) 'AI Skills Improve Job Prospects: Causal Evidence from a Hiring Experiment'. arXiv. Available at: https://doi.org/10.48550/arXiv.2601.13286.

Stöhr, C., Ou, A.W. and Malmström, H. (2024) 'Perceptions and usage of AI chatbots among students in higher education across genders, academic levels and fields of study', *Computers and Education: Artificial Intelligence*, 7, p. 100259. Available at: https://doi.org/10.1016/j.caeai.2024.100259.

Tang, C. *et al.* (2025) 'Gender disparities in the impact of generative artificial intelligence: Evidence from academia', *PNAS Nexus*, 4(2), p. pgae591. Available at: https://doi.org/10.1093/pnasnexus/pgae591.

Thébaud, S. and Charles, M. (2018) 'Segregation, Stereotypes, and STEM', *Social Sciences*, 7(7), p. 111. Available at: https://doi.org/10.3390/socsci7070111.

Tifferet, S. (2019) 'Gender differences in privacy tendencies on social network sites: A meta-analysis', *Computers in Human Behavior*, 93, pp. 1–12. Available at: https://doi.org/10.1016/j.chb.2018.11.046.

Toft, M.B., Schuitema, G. and Thøgersen, J. (2014) 'Responsible technology acceptance: Model development and application to consumer acceptance of Smart Grid technology', *Applied Energy*, 134, pp. 392–400. Available at: https://doi.org/10.1016/j.apenergy.2014.08.048.

Treuthart, M.P. (2019) 'Connectivity: The Global Gender Digital Divide and Its Implications for Women's Human Rights and Equality', *Gonzaga Journal of International Law*, 23(1), pp. 1–54.

Wang, M.-T. and Degol, J.L. (2017) 'Gender Gap in Science, Technology, Engineering, and Mathematics (STEM): Current Knowledge, Implications for Practice, Policy, and Future Directions', *Educational Psychology Review*, 29(1), pp. 119–140. Available at: https://doi.org/10.1007/s10648-015-9355-x.

Yoshikawa, K., Kokubo, A. and Wu, C.-H. (2018) 'A Cultural Perspective on Gender Inequity in STEM: The Japanese Context', *Industrial and Organizational Psychology*, 11(2), pp. 301–309. Available at: https://doi.org/10.1017/iop.2018.19.

Young, E., Wajcman, J. and Sprejer, L. (2023) 'Mind the gender gap: Inequalities in the emergent professions of artificial intelligence (AI) and data science', *New Technology, Work and Employment*, 38(3), pp. 391–414. Available at: https://doi.org/10.1111/ntwe.12278.

Yufei, M., Xiangquan, Z. and Wenxin, H. (2018) 'Does Internet Usage Reduce the Gender Wage Gap? Empirical Analysis based on CFPS Data', *Journal of Finance and Economics*, 44(07), pp. 33–45. Available at: https://doi.org/10.16538/j.cnki.jfe.2018.07.003.

Zander, L. *et al.* (2020) 'When Grades Are High but Self-Efficacy Is Low: Unpacking the Confidence Gap Between Girls and Boys in Mathematics', *Frontiers in Psychology*, 11. Available at: https://doi.org/10.3389/fpsyg.2020.552355.

# Appendix

**Table 1**: Sample Characteristics (Weighted, 2023 vs 2024)

| | 2023 | | 2024 | |
|---|---|---|---|---|
| | Sample Size | (%) | Sample Size | (%) |
| **Gender** | | | | |
| Female | 2173 | 51% | 2549 | 52% |
| Male | 2027 | 48% | 2378 | 48% |
| Identify in another way | 15 | <1% | 15 | <1% |
| Prefer not to say | 10 | <1% | 5 | <1% |
| **Age** | | | | |
| NET: 18–34 | 1175 | 28% | 1364 | 28% |
| NET: 35–54 | 1391 | 33% | 1622 | 33% |
| NET: 55+ | 1659 | 39% | 1961 | 40% |
| **Socio-economic classification** | | | | |
| ABC1 | 2235 | 53% | 2746 | 56% |
| C2DE | 1990 | 47% | 2201 | 45% |
| **Region** | | | | |
| Northern Ireland | 116 | 3% | 137 | 3% |
| Scotland | 351 | 8% | 409 | 8% |
| North-West | 465 | 11% | 547 | 11% |
| North-East | 167 | 4% | 199 | 4% |
| Yorkshire & Humberside | 342 | 8% | 404 | 8% |
| Wales | 199 | 5% | 231 | 5% |
| West Midlands | 374 | 9% | 435 | 9% |
| East Midlands | 307 | 7% | 361 | 7% |
| South-West | 366 | 9% | 427 | 9% |
| South-East | 581 | 14% | 681 | 14% |
| Eastern | 396 | 9% | 460 | 9% |
| London | 561 | 13% | 656 | 13% |
| NET England | 3559 | 84% | 4170 | 84% |

**Table 2:** Gender differences in frequent personal GenAI use, ordered by gender gap

| Feature | Frequent use by Women (%) | Frequent use by Men (%) | Gender Gap (pp) | N | Women in Group (%) | Frequent Use (All, %) |
|---|---|---|---|---|---|---|
| AI risk: mental health | 13.92 | 33.41 | 19.49 | 490 | 56.76 | 22.21 |
| Age 35–44 | 16.42 | 29.23 | 12.81 | 850 | 53.28 | 22.33 |
| Age 25–34 | 25.85 | 38.29 | 12.44 | 862 | 54.57 | 31.39 |
| Age 18–24 | 23.71 | 33.11 | 9.40 | 520 | 70.28 | 26.63 |
| AI risk: data privacy | 11.77 | 20.98 | 9.22 | 1205 | 49.24 | 16.44 |
| AI risk: climate impact | 18.71 | 27.77 | 9.06 | 426 | 47.74 | 23.12 |
| Concern: no choice about data sharing | 12.76 | 20.71 | 7.95 | 1579 | 52.61 | 16.56 |
| Digital literacy: can explain AI in detail | 34.36 | 42.06 | 7.70 | 761 | 43.34 | 38.50 |
| Concern: data leaves people behind | 14.57 | 22.12 | 7.55 | 1164 | 50.62 | 18.33 |
| Concern: data sold to third parties | 11.24 | 18.34 | 7.10 | 2700 | 51.72 | 14.61 |
| AI risk: job displacement | 11.88 | 18.74 | 6.86 | 1864 | 54.02 | 14.99 |
| Feels safe from cyber attacks | 26.75 | 33.53 | 6.78 | 1469 | 46.10 | 30.42 |
| Age 45–54 | 13.14 | 19.90 | 6.76 | 814 | 50.28 | 16.48 |
| AI risk: bias | 15.83 | 22.20 | 6.36 | 676 | 45.79 | 19.23 |
| Uses GenAI at work | 60.22 | 66.53 | 6.30 | 696 | 47.89 | 63.67 |
| Believes AI is a big societal risk | 15.22 | 21.24 | 6.02 | 232 | 54.45 | 17.89 |
| University degree or higher | 20.31 | 26.23 | 5.92 | 2227 | 51.81 | 23.15 |
| All respondents (baseline) | 14.70 | 20.44 | 5.74 | 4927 | 51.74 | 17.50 |
| Managerial/clerical (AB/C1) | 15.42 | 21.14 | 5.71 | 2809 | 51.65 | 18.17 |
| Digital literacy: can explain AI | 17.68 | 23.23 | 5.55 | 3501 | 50.27 | 20.41 |
| Age 75+ | 1.51 | 6.83 | 5.32 | 377 | 42.75 | 4.55 |
| AI risk: loss of human agency | 10.60 | 15.91 | 5.31 | 1400 | 50.54 | 13.22 |
| Concern: data not held securely | 11.79 | 17.05 | 5.26 | 2880 | 54.04 | 14.19 |
| Age 55–64 | 6.91 | 12.10 | 5.19 | 868 | 46.87 | 9.65 |
| Higher/intermediate managerial | 18.08 | 23.25 | 5.17 | 1219 | 50.18 | 20.65 |
| Concern: data harms environment | 27.53 | 32.66 | 5.13 | 547 | 50.36 | 30.39 |
| Digital literacy: AI training knowledge | 37.43 | 42.36 | 4.93 | 1257 | 45.30 | 40.16 |
| Societal AI optimism | 28.68 | 33.54 | 4.86 | 1944 | 46.19 | 31.32 |
| Personal AI optimism | 28.68 | 33.54 | 4.86 | 1944 | 46.19 | 31.32 |
| AI risk: terrorism | 13.33 | 18.04 | 4.71 | 965 | 55.50 | 15.47 |
| Concern: data used for surveillance | 14.47 | 19.07 | 4.60 | 1628 | 45.04 | 16.98 |
| AI risk: unexplainability | 15.45 | 19.73 | 4.28 | 903 | 52.86 | 17.50 |
| AI risk: creativity erosion | 12.56 | 16.27 | 3.71 | 1496 | 57.48 | 14.13 |
| AI risk: disinformation | 12.38 | 15.99 | 3.61 | 1402 | 50.42 | 14.37 |
| Age 65–74 | 5.90 | 8.13 | 2.24 | 656 | 46.27 | 7.40 |
| AI risk: accountability | 15.18 | 17.07 | 1.89 | 1027 | 49.72 | 16.08 |

Table 3: Gender differences in frequent work GenAI use, ordered by gender gap

| Feature | Frequent use by Women (%) | Frequent use by Men (%) | Gender Gap (pp) | N | Women in Group (%) | Frequent Use (All, %) |
|---|---|---|---|---|---|---|
| AI risk: mental health | 11.55 | 23.52 | 11.97 | 490 | 56.76 | 16.62 |
| Age 18–24 | 20.04 | 27.99 | 7.94 | 520 | 70.28 | 22.57 |
| Age 35–44 | 16.52 | 24.19 | 7.67 | 850 | 53.28 | 20.06 |
| AI risk: data privacy | 9.52 | 15.67 | 6.15 | 1205 | 49.24 | 12.58 |
| Age 45–54 | 8.86 | 14.73 | 5.87 | 814 | 50.28 | 11.77 |
| Digital literacy: can explain AI in detail | 30.00 | 35.80 | 5.80 | 761 | 43.34 | 33.11 |
| Believes AI is a big societal risk | 10.55 | 15.52 | 4.98 | 232 | 54.45 | 12.76 |
| Age 25–34 | 28.16 | 32.39 | 4.22 | 862 | 54.57 | 29.97 |
| Concern: no choice about data sharing | 12.05 | 16.24 | 4.19 | 1579 | 52.61 | 14.02 |
| Digital literacy: AI training knowledge | 32.68 | 36.39 | 3.71 | 1257 | 45.30 | 34.62 |
| Managerial/clerical (AB/C1) | 14.62 | 17.97 | 3.36 | 2809 | 51.65 | 16.20 |
| Concern: data harms environment | 24.37 | 27.19 | 2.82 | 547 | 50.36 | 25.77 |
| AI risk: accountability | 12.11 | 14.86 | 2.75 | 1027 | 49.72 | 13.45 |
| AI risk: climate impact | 20.25 | 22.75 | 2.50 | 426 | 47.74 | 21.25 |
| AI risk: job displacement | 10.38 | 12.70 | 2.32 | 1864 | 54.02 | 11.41 |
| Feels safe from cyber attacks | 24.01 | 26.25 | 2.24 | 1469 | 46.10 | 25.18 |
| Age 55–64 | 4.15 | 6.37 | 2.22 | 868 | 46.87 | 5.32 |
| All respondents (baseline) | 13.00 | 15.16 | 2.17 | 4927 | 51.74 | 14.05 |
| University degree or higher | 19.82 | 21.93 | 2.11 | 2227 | 51.81 | 20.83 |
| Digital literacy: can explain AI | 15.29 | 17.26 | 1.97 | 3501 | 50.27 | 16.22 |
| Concern: data sold to third parties | 9.78 | 11.65 | 1.87 | 2700 | 51.72 | 10.64 |
| Concern: data not held securely | 9.28 | 11.12 | 1.84 | 2880 | 54.04 | 10.09 |
| Age 75+ | 0.50 | 2.32 | 1.82 | 377 | 42.75 | 1.54 |
| Concern: data leaves people behind | 14.94 | 16.65 | 1.71 | 1164 | 50.62 | 15.83 |
| Concern: data used for surveillance | 10.74 | 12.07 | 1.34 | 1628 | 45.04 | 11.48 |
| Societal AI optimism | 25.53 | 26.54 | 1.01 | 1944 | 46.19 | 26.07 |
| Personal AI optimism | 25.53 | 26.54 | 1.01 | 1944 | 46.19 | 26.07 |
| Age 65–74 | 1.99 | 2.96 | 0.97 | 656 | 46.27 | 2.51 |
| Higher/intermediate managerial | 19.11 | 20.00 | 0.89 | 1219 | 50.18 | 19.47 |
| AI risk: loss of human agency | 10.51 | 10.99 | 0.48 | 1400 | 50.54 | 10.74 |
| AI risk: terrorism | 12.30 | 12.36 | 0.07 | 965 | 55.50 | 12.28 |
| Uses GenAI at work | 100.00 | 100.00 | 0.00 | 696 | 47.89 | 100.00 |
| AI risk: creativity erosion | 12.41 | 12.16 | -0.24 | 1496 | 57.48 | 12.31 |
| AI risk: unexplainability | 14.79 | 14.46 | -0.33 | 903 | 52.86 | 14.68 |
| AI risk: bias | 14.43 | 14.09 | -0.34 | 676 | 45.79 | 14.20 |
| AI risk: disinformation | 10.34 | 9.82 | -0.52 | 1402 | 50.42 | 10.24 |

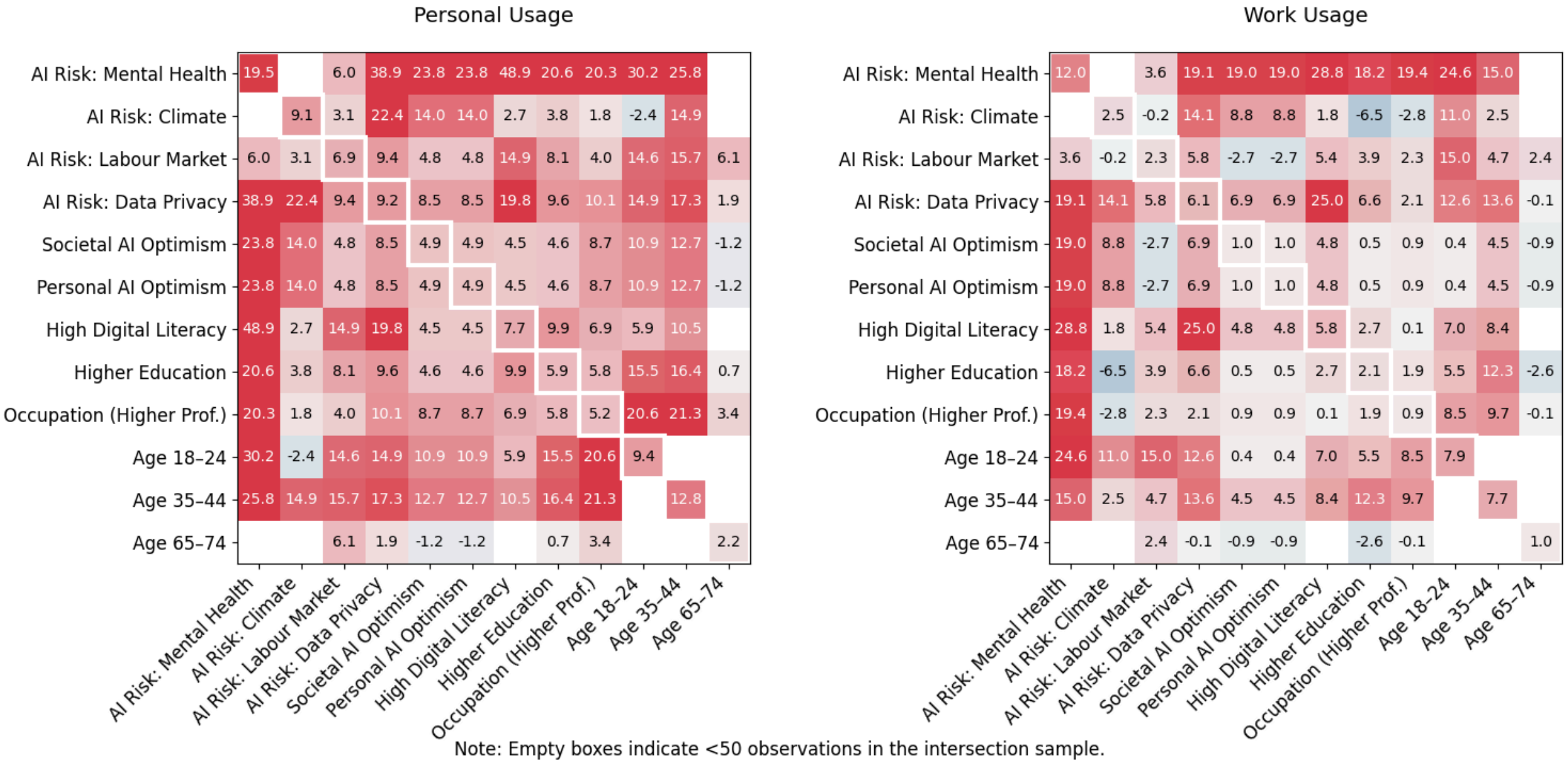

**Figure A1. Intersectional gender gaps in GenAI usage across paired factors and attitudes.** The heatmaps show gender differences (percentage-point gaps) in frequent personal (left) and work (right) use of generative AI for respondents when two characteristics or attitudes intersect. Rows represent the primary factor of interest; columns represent secondary intersecting factors. Positive values indicate higher usage among men. Empty cells reflect intersections with fewer than 50 observations.

Figure A1 maps how gender gaps in GenAI adoption widen or narrow when two factors or attitudes intersect. Consistent with the main findings, gender gaps in work-related use are generally smaller than those for personal use across most intersections. A notable exception concerns respondents who view AI as a risk to mental health: for this group, work-related gaps remain similar to personal use gaps, despite the lower overall baseline in the work context (12.0 vs 19.5 percentage points, respectively). The single largest intersectional gap emerges among individuals who both perceive AI as a mental health risk and report high digital literacy (48.9 percentage points in personal use vs 28.8 in work use). Younger adults (18–24) consistently display wide gaps in both contexts. Among those concerned about mental health harms, most additional factors widen the disparity—including privacy concerns (38.9pp) and strong AI fluency (48.9pp)—with one notable moderating exception: pairing mental health concern with labour market concern yields much smaller gaps (6.0pp personal; 3.6pp work). In the work context, gender gaps rise sharply for young respondents who also hold additional worries about AI (e.g., AI as a mental health threat: 24.6pp; labour market threat: 15.0pp; climate harm: 11.0pp), but shrink to near zero for those who are optimistic about AI's societal or personal impact. Privacy concerns coupled with high digital literacy emerges as an outlier in that the gap is wider in the work context than in personal use, respectively at 25.0pp and 19.8pp. Overall, intersectional attitudes meaningfully modulate the gender gap, highlighting how concern domains and digital fluency jointly shape uneven GenAI uptake.

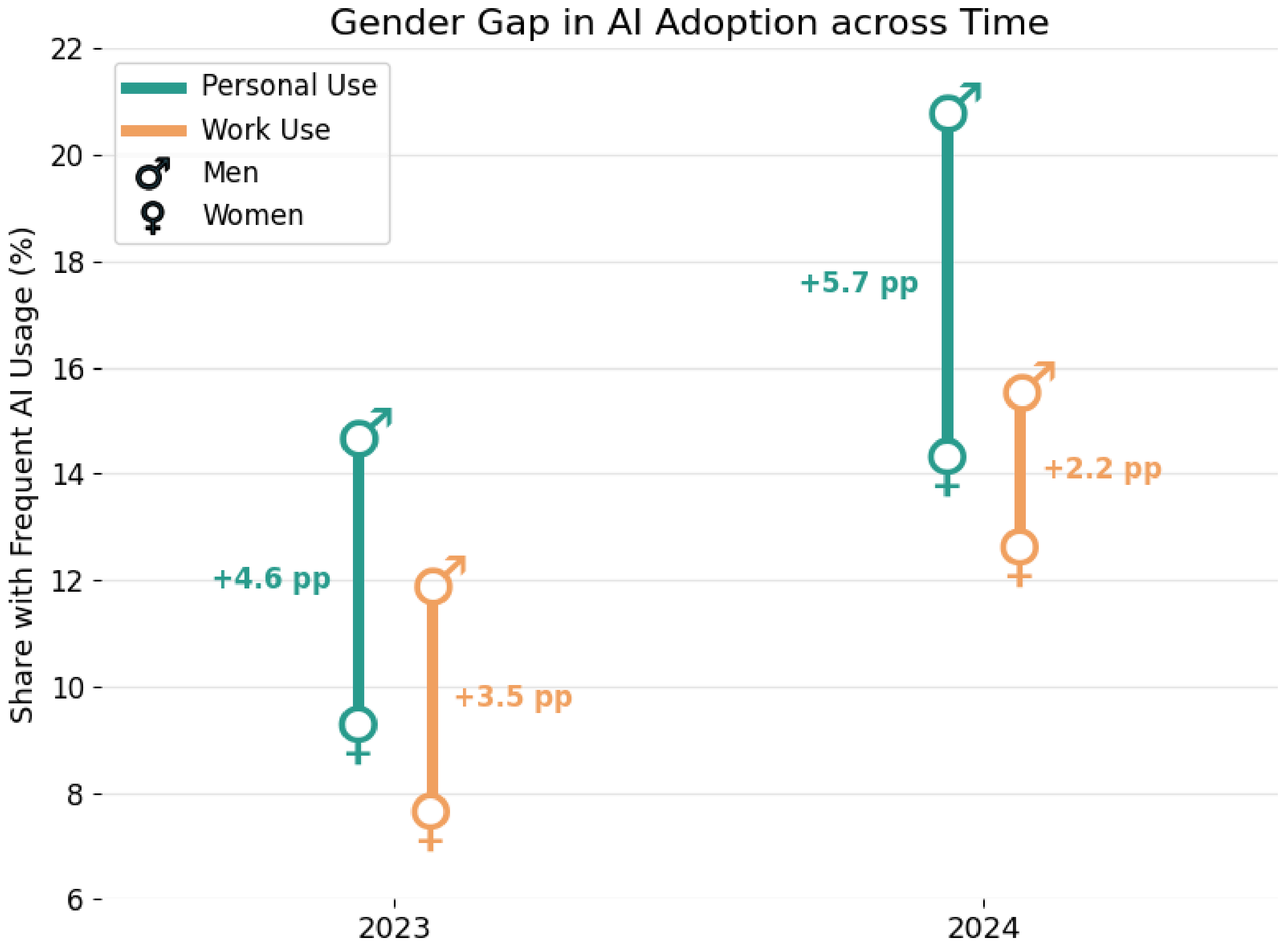

**Figure A2. Gender gaps in frequent GenAI use between 2023 and 2024.**
Male–female differences in frequent personal and work-related use of generative AI across two survey waves (2023 and 2024). Markers show usage rates for men and women each year. Personal use gaps widen from +4.6 pp in 2023 to +5.7 pp in 2024, whereas work use gaps narrow from +3.5 pp to +2.2 pp.

Figure A2 shows how gender disparities in GenAI adoption have shifted over time. Between 2023 and 2024, frequent personal use rose substantially for both men (14.3% to 20.4%) and women (9.7% to 14.7%), but men's faster uptake led to a widening gender gap in personal use—from 4.6pp to 5.7pp. In contrast, work-related adoption increased more modestly (men: 11.5% to 15.2%; women: 8.0% to 13.0%), producing a narrowing of the work-use gap from 3.5 pp to 2.2 pp. These patterns suggest that while overall usage is rising, personal adoption is accelerating more quickly among men—reinforcing inequality—whereas workplace adoption appears to be converging modestly across genders.

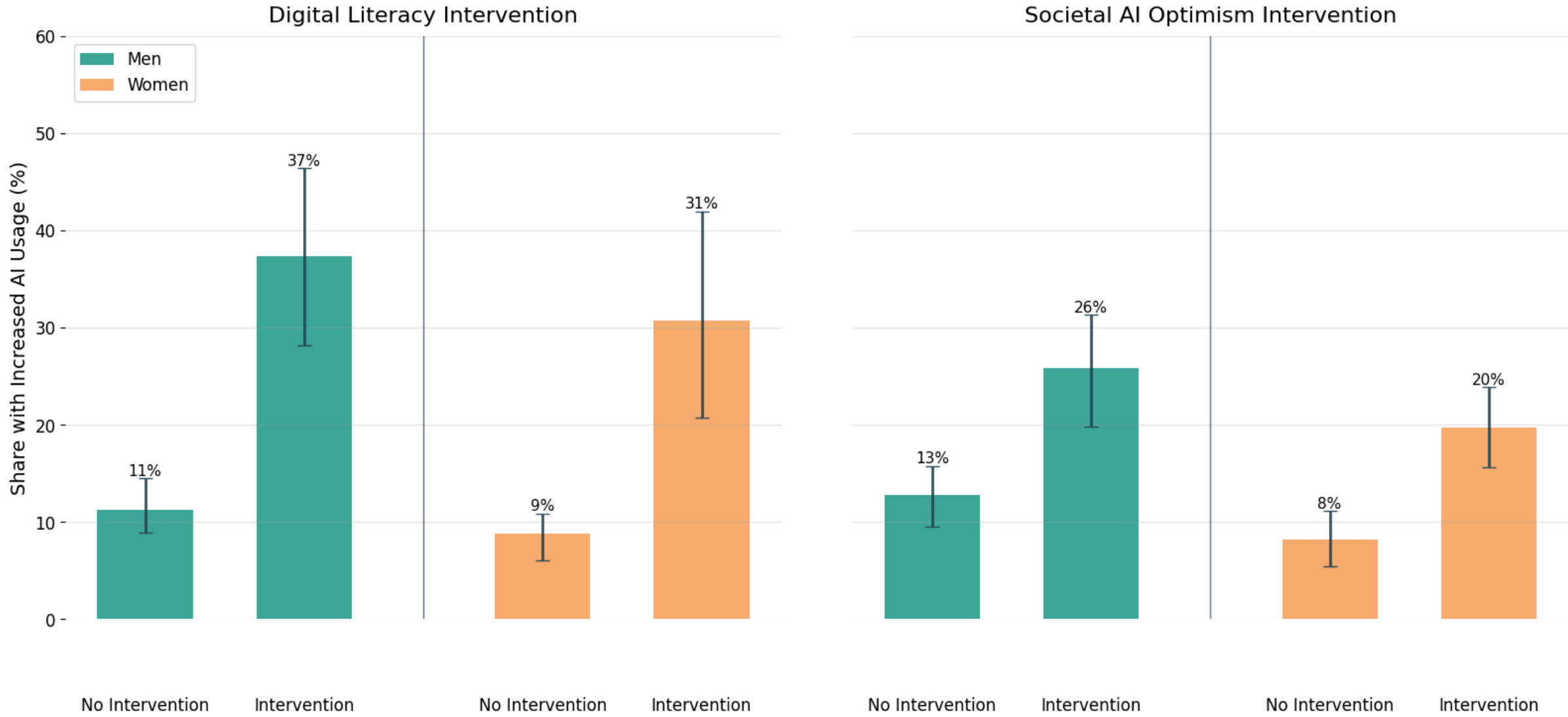

**Figure A3. Digital literacy, societal AI optimism, and matched differences in personal GenAI use (full sample).**

Score-matched comparisons of higher digital literacy (left) and higher optimism about AI's societal impact (right) with personal GenAI uptake between survey waves (2023–2024). Estimates are based on 100 iterations of a parametric score-matching procedure pairing Wave 3 respondents with demographically comparable Wave 4 respondents, matched on age, gender, education, and occupation. Error bars represent 95% confidence intervals.

For digital literacy, respondents were classified on a four-point scale of understanding of generative AI: no understanding, little understanding, partial understanding, and full understanding. Within matched comparisons, the "higher literacy" group consists of cases where the Wave 4 respondent reports a higher level of GenAI understanding than the matched Wave 3 respondent. The comparison group consists of matched cases where the Wave 4 respondent does not report higher literacy. We then compare the share of matched cases in which personal GenAI use is higher in 2024 than among comparable respondents in 2023. Analyses are conducted separately for men and women. An analogous procedure is applied to societal AI optimism, comparing matched cases where the Wave 4 respondent reports higher optimism with cases where optimism is not higher.

Figure A3 shows the full-sample score-matched comparisons, which differ somewhat from the young-adult results reported in the main text. In the full sample, both higher digital literacy and higher societal AI optimism are associated with greater GenAI uptake among men and women. The association between higher digital literacy and uptake remains stronger for men than for women, echoing the young-adult pattern. However, unlike among younger adults, women in the full sample also show a clear positive association between higher literacy and increased GenAI use. For societal AI optimism, the association is positive for both men and women, and slightly stronger for women.